\documentclass[conference]{IEEEtran}
\IEEEoverridecommandlockouts
\usepackage{cite}
\usepackage{amsmath,amssymb,amsfonts}
\usepackage{algorithmic}
\usepackage{graphicx}
\usepackage{textcomp}
\usepackage{xcolor}
\usepackage{url}
\usepackage{multirow}
\usepackage{makecell}
\usepackage[most]{tcolorbox}
\usepackage{enumitem}

\newcommand{\hdashline}{%
  \noindent
  \begin{tikzpicture}
    \draw[dashed,gray] (0,0) -- (\linewidth,0);
  \end{tikzpicture}%
}

\tcbset{
  boxrule=0.3pt,
  arc=6pt,
  left=3pt, right=3pt, top=3pt, bottom=3pt,
  boxsep=1.5pt,
}

\newtcolorbox{userbox}{
  width=0.9\columnwidth,
  colback=white,
  colframe=black!30,
  arc=6pt,
  fontupper=\scriptsize,
}

\newtcolorbox{llmbox}{
  width=0.9\columnwidth,
  colback=white,
  colframe=black!30,
  arc=6pt,
  fontupper=\scriptsize,
}

\newtcolorbox{toolbox}{
  width=0.9\columnwidth,
  colback=gray!15,
  colframe=black!15,
  arc=6pt,
  fontupper=\scriptsize,
  left=0.3mm, right=0.3mm, boxsep=0.2mm,
  before skip=0pt, after skip=0pt,
}

\newtcolorbox{obsboxenv}{
  width=0.9\columnwidth,
  colback=gray!15,
  colframe=black!15,
  arc=6pt,
  fontupper=\scriptsize\ttfamily,
  listing only,
  listing options={
    basicstyle=\scriptsize\ttfamily,
    breaklines=true,
    columns=fullflexible,
  },
}

\newenvironment{observation}
 {\begin{flushright}\begin{obsboxenv}{\normalfont\scriptsize\bfseries Observation:}\\}
 {\end{obsboxenv}\end{flushright}}

\newtcolorbox{finalbox}{
  width=\linewidth,
  colback=gray!25,
  colframe=black!30,
  arc=6pt,
  fontupper=\scriptsize,
  left=3pt, right=3pt, top=3pt, bottom=3pt,
  boxsep=1.5pt,
}

\newcommand{\userturn}[1]{%
  \begin{flushleft}
    \begin{userbox}
      \textbf{Question:}\\[0.25ex]
      #1
    \end{userbox}
  \end{flushleft}
}

\newcommand{\tool}[1]{%
  \begin{flushleft}
    \begin{toolbox}
      \textbf{Tool:} #1
    \end{toolbox}
  \end{flushleft}
}

\newcommand{\finalanswer}[1]{%
  \begin{finalbox}
    \textbf{Answer:}\\[0.25ex]
    #1
  \end{finalbox}
}

\def\BibTeX{{\rm B\kern-.05em{\sc i\kern-.025em b}\kern-.08em
    T\kern-.1667em\lower.7ex\hbox{E}\kern-.125emX}}
\begin{document}

\title{Querying Large Automotive Software Models: Agentic vs. Direct LLM Approaches \\
\thanks{
This work has received funding from the European Chips Joint Undertaking under Framework Partnership Agreement No 101139789 (HAL4SDV) including the national funding from the German Federal Ministry of Education and Research (BMBF) under grant number 16MEE00471K. The responsibility for the content of this publication lies with the authors.
}
}

\author{
\IEEEauthorblockN{Lukasz Mazur}
\IEEEauthorblockA{
\textit{Chair of Robotics, Artificial} \\ 
\textit{Intelligence and Real-Time Systems} \\
\textit{Technical University of Munich}\\
Munich, Germany \\
lukasz.mazur@tum.de}
\and
\IEEEauthorblockN{Nenad Petrovic}
\IEEEauthorblockA{
\textit{Chair of Robotics, Artificial} \\ 
\textit{Intelligence and Real-Time Systems} \\
\textit{Technical University of Munich}\\
Munich, Germany \\
nenad.petrovic@tum.de}
\and
\IEEEauthorblockN{James Pontes Miranda}
\IEEEauthorblockA{
\textit{Software and Systems Engineering} \\
\textit{Université Paris-Saclay} \\
\textit{CEA List} \\
Palasieu, France \\
james.pontesmiranda@cea.fr}
\and
\IEEEauthorblockN{Ansgar Radermacher}
\IEEEauthorblockA{
\textit{Software and Systems Engineering} \\
\textit{Université Paris-Saclay} \\
\textit{CEA List} \\
Palaiseau, France \\
ansgar.radermacher@cea.fr}
\and
\IEEEauthorblockN{Robert Rasche}
\IEEEauthorblockA{
\textit{Real-Time Systems, Resources,} \\
\textit{Model Based Software Eng.} \\
\textit{Tensor embedded GmbH} \\
Pollenfeld, Germany \\
robert.rasche@tensor.gmbh}
\and
\IEEEauthorblockN{Alois Knoll}
\IEEEauthorblockA{
\textit{Chair of Robotics, Artificial} \\ 
\textit{Intelligence and Real-Time Systems} \\
\textit{Technical University of Munich}\\
Munich, Germany \\
knoll@in.tum.de}
}

\maketitle

\begin{abstract}
Large language models (LLMs) offer new opportunities for interacting with complex software artifacts, such as software models, through natural language. They present especially promising benefits for large software models that are difficult to grasp in their entirety, making traditional interaction and analysis approaches challenging. This paper investigates two approaches for leveraging LLMs to answer questions over software models: direct prompting, where the whole software model is provided in the context, and an agentic approach combining LLM-based agents with general-purpose file access tools. We evaluate these approaches using an Ecore metamodel designed for timing analysis and software optimization in automotive and embedded domains. Our findings show that while the agentic approach achieves accuracy comparable to direct prompting, it is significantly more efficient in terms of token usage. This efficiency makes the agentic approach particularly suitable for the automotive industry, where the large size of software models makes direct prompting infeasible, establishing LLM agents as not just a practical alternative but the only viable solution. Notably, the evaluation was conducted using small LLMs, which are more feasible to be executed locally — an essential advantage for meeting strict requirements around privacy, intellectual property protection, and regulatory compliance. Future work will investigate software models in diverse formats, explore more complex agent architectures, and extend agentic workflows to support not only querying but also modification of software models.
\end{abstract}


\section{Introduction}

Software modeling is crucial in modern software engineering, particularly in safety-critical domains like automotive development, where precise models ensure compliance and reliability~\cite{broy_engineering_2007}. However, the increasing size and complexity of these models pose significant challenges for engineers in terms of access, understanding, and manipulation~\cite{haghighatkhah_automotive_2017}.

Recent advancements in large language models (LLMs) offer promising avenues to assist with complex software models, leveraging their capabilities in natural language understanding and code generation~\cite{fan_large_2023}. Despite this potential, the direct application of LLMs to structured software models, especially with formal languages like PlantUML~\cite{noauthor_plantuml_nodate}, remains largely unexplored~\cite{muff_limitations_2024}. Current research primarily focuses on unstructured text or code, overlooking the unique challenges of integrating LLMs with structured models, including format compatibility, semantic precision, and consistency. Furthermore, the efficacy of different LLM architectures, such as agentic workflows versus direct prompting, in understanding and reasoning about software models is unclear.

This study investigates and compares agentic LLM workflows and direct full-context prompting for querying software models. We evaluate these approaches on specific, concrete questions about model structure and semantics, using a representative automotive software model.

Our contributions are twofold: an empirical comparison of LLM-based agentic and direct prompting for software model question answering, and insights into the feasibility and limitations of current LLM capabilities in this context. These findings will inform future research on advanced interactions with software models in model-driven engineering.

The paper is structured as follows: Section 2 reviews related work. Section 3 details the methodology. Section 4 presents results and analysis. Section 5 discusses findings. Section 6 addresses threats to validity, and Section 7 concludes with future work.

\section{Related Work}

\subsection{LLMs in Software and Model-Driven Engineering}

LLMs have increasingly been applied in various software engineering tasks, including code generation, bug detection, documentation, and summarization. Their ability to understand and generate natural language alongside programming languages makes them valuable assistants for developers across the software development lifecycle~\cite{fan_large_2023}.

In model-driven engineering (MDE), LLMs are increasingly being explored for tasks such as model generation, validation, and transformation, aiming to improve automation and reduce manual effort~\cite{fan_large_2023}. While software models offer high-level abstractions of system designs, their structured and formal nature poses challenges for natural language models, requiring specialized input representations and reasoning strategies~\cite{dirocco_use_2025}. A variety of modeling standards are used across domains, including UML~\cite{noauthor_unified_nodate} and Eclipse Modeling Framework (EMF)~\cite{noauthor_eclipse_nodate} in general software engineering, and AUTOSAR~\cite{noauthor_autosar_nodate} in the automotive industry. These standards provide precise specifications of structure and behavior, which are critical in safety-critical domains governed by standards like ISO 26262, highlighting the growing need to integrate AI-assisted techniques into rigorous development processes~\cite{vinoth_kannan_model-based_2021}.

Traditional software modeling tools provide querying, validation, and transformation capabilities to manage these complex models. EMF and specialized automotive toolchains support model editing and consistency checks~\cite{vinoth_kannan_model-based_2021}\cite{mosquera_understanding_2024}. The integration of LLMs into such environments has the potential to enhance these tools by enabling natural language querying, automated documentation generation, and intelligent assistance during modeling. This intersection of LLM technology with established software modeling ecosystems remains an active area of exploration~\cite{dirocco_use_2025}.

\subsection{LLM Integration with Structured Software Artifacts}

Interfacing LLMs with structured software artifacts such as models, codebases, or graphs requires specialized architectural approaches. Unlike plain text, software models are inherently structured, hierarchical, and often constrained by formal metamodels, which challenges traditional LLM input paradigms~\cite{sui_table_2024}.

One prominent technique to address these challenges is Retrieval-Augmented Generation (RAG), where external knowledge bases are queried to retrieve relevant fragments that augment the LLM’s context. This approach helps mitigate token-length constraints and context fragmentation, enabling the LLM to focus on pertinent information without being overwhelmed by the entire knowledge base~\cite{gao_retrieval-augmented_2024}.

Beyond retrieval-based methods, tool-augmented LLM architectures enable more sophisticated workflows. These agents can perform iterative reasoning, plan multi-step queries, and invoke external tools or APIs for accessing external data sources. Agentic approaches combine LLM capabilities with domain-specific logic and memory management, leading to improved performance on complex tasks~\cite{sumers_cognitive_2023}\cite{paranjape_art_2023}. As tool usage gains popularity, emerging standards such as the Model Context Protocol (MCP)~\cite{noauthor_model_nodate} and Agent2Agent Protocol (A2A)~\cite{noauthor_agent2agent_nodate} are being developed to support scalable and interoperable agent ecosystems. MCP standardizes how agents connect to tools, APIs, and resources through structured inputs and outputs, while A2A facilitates dynamic, multimodal communication between agents, enabling collaboration, delegation, and shared task management.

The balance between single-shot prompting and multi-turn agentic workflows also influences system capabilities. While single-shot prompts are simpler and faster, agentic workflows provide enhanced reasoning, planning, and verification, but come with higher computational costs and increased latency~\cite{wei_chain--thought_2022}\cite{wang_self-consistency_2022}. The choice between these architectures typically depends on the task's complexity and how well the agent's available tools align with the task's abstraction level.

\subsection{Evaluation of Free-Text LLM Answers}

Evaluating open-ended natural language answers from LLMs is challenging, especially when accuracy, semantic alignment, and factual grounding are essential. Traditional lexical matching metrics that focus on the overlap of n-grams (word sequences) between two texts, like BLEU and ROUGE, often fall short when evaluating the quality of responses to structured or knowledge-intensive queries because they primarily assess surface-level word sequence commonality~\cite{kamalloo_evaluating_2023}. While semantic similarity methods, such as bi-encoder and cross-encoder models, better capture meaning beyond lexical overlap, they still struggle with evaluating factual accuracy, logical consistency, and adherence to task-specific constraints in complex scenarios~\cite{mustafazade_evaluation_2022}.

LLM-as-a-Judge methods have recently emerged as a flexible and scalable evaluation paradigm, prompting LLMs to assess the correctness, completeness, and relevance of generated answers using their own domain understanding. However, these models may still hallucinate or misjudge, limiting their reliability~\cite{wang_can_2025}. Human evaluation remains essential for ensuring validity, particularly for complex reasoning tasks, but it is resource-intensive and often requires expert reviewers to assess nuanced correctness, adherence to software engineering principles, and practical relevance~\cite{badshah_reference-guided_2024}.

\section{Methodology}

\subsection{Research Questions}

This study investigates how LLMs and agent-based architectures perform on question-answering tasks over complex software models, comparing direct full-context prompting with agentic retrieval using external tools.

\textit{RQ1: How does the accuracy of an LLM-based agent that retrieves and processes parts of a large software model using external tools compare to the accuracy of a reference LLM that processes the entire model within its context window when answering model-related questions?}

This question targets the core trade-off between full-context access and modular, tool-assisted access. It aims to evaluate whether an LLM-based agent can still produce accurate and useful answers to domain-relevant questions despite never having a complete view of the software model at once. The answer to this question has significant implications for the feasibility of using LLM agents in real-world automotive workflows where models may be large, distributed, and subject to change.

\textit{RQ2: How does the token consumption of an LLM-based agent that incrementally accesses a large software model via tool support compare to that of an LLM processing the entire model in a single pass?}

This question addresses the efficiency of the two approaches in terms of token usage, which is especially critical when deploying smaller LLMs with limited context windows. Smaller models are often the only viable option for local, on-premise deployment - an increasingly important requirement in the automotive industry due to strict privacy constraints, intellectual property protection, and regulatory compliance. By analyzing token consumption, we aim to understand whether agent-based strategies enable smaller, locally hosted models to work effectively with large-scale software artifacts.

\subsection{Direct Prompting and Agent-Based Architecture}

\begin{figure*}[htbp]
\centerline{\includegraphics[width=\textwidth]{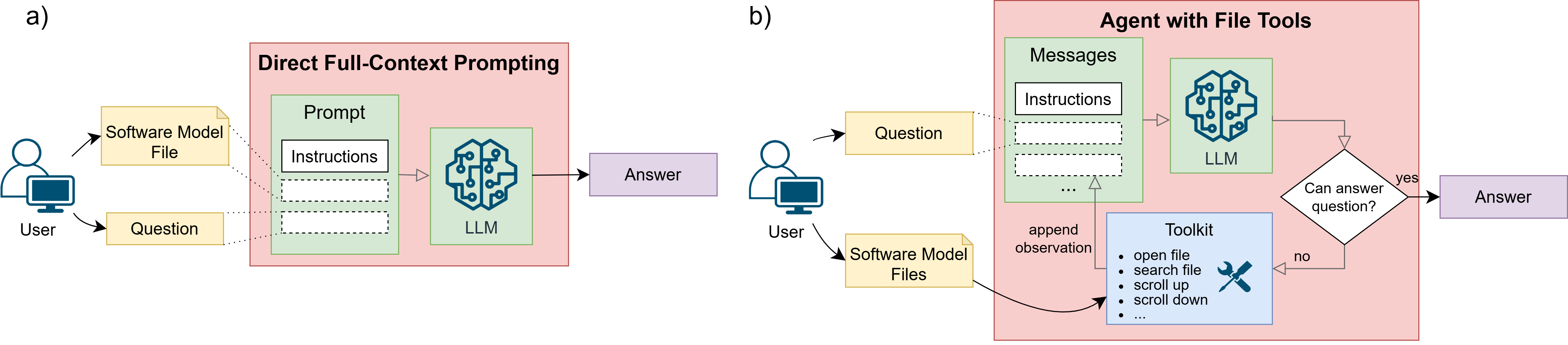}}
\caption{Evaluation setups for software model question answering. (a) Direct LLM call with full model in prompt. (b) Agent using file-access tools to retrieve relevant model parts before answering.}
\label{fig_architectures}
\end{figure*}

To evaluate the ability of LLMs to answer questions about complex software models, we compare two architectures: a direct LLM call serving as a reference, and a tool-augmented agent architecture. Both setups are illustrated in Fig.~\ref{fig_architectures}, with example responses shown in Fig.~\ref{fig_example_responses} to demonstrate how each approach processes and answers a question.

\begin{figure}[t]
\centering
\begingroup
\setlength{\parskip}{0pt}

\begin{minipage}{\columnwidth}
  \userturn{What are the fields of Frequency class? Provide names and data types of only this class, don't list fields of base classes.}
  \vspace{-0.7em}
  \hdashline
\end{minipage}

\vspace{0.4em}

\noindent
\begin{tabular}[t]{@{} >{\raggedright\arraybackslash}p{1.5em} @{\hspace{0.5em}} p{0.84\columnwidth} @{}}
\raisebox{7em}[0pt][0pt]{\textbf{(a)}} &
\begin{minipage}[t]{\linewidth}
  \finalanswer{
  The Frequency class has the following fields defined directly:

  1. name: value \\
  \hspace*{0.5em} - data type: EDouble (double) \\
  \hspace*{0.5em} - lowerBound: 1 \\
  \hspace*{0.5em} - defaultValueLiteral: 1000

  2. name: unit \\
  \hspace*{0.5em} - data type: FrequencyUnit (enumeration) \\
  \hspace*{0.5em} - lowerBound: 1 \\
  \hspace*{0.5em} - defaultValueLiteral: MHz
  }
\end{minipage}
\end{tabular}

\vspace{-0.5em}
\hdashline
\vspace{0.5em}

\noindent
\begin{tabular}[t]{@{} >{\raggedright\arraybackslash}p{1.5em} @{\hspace{0.5em}} p{0.84\columnwidth} @{}}
\raisebox{0pt}[0pt][0pt]{\textbf{(b)}} &
\begin{minipage}[t]{\linewidth}
  \tool{\texttt{open\_file root.ecore}}
  \begin{observation}
  ... first 50 lines of the file ...
  \end{observation}

  \tool{\texttt{search\_file Frequency root.ecore}}
  \begin{observation}
  Found 8 matches for "Frequency" in root.ecore:\\
  ...\\
  Line 2830: name="Frequency">\\
  ...
  \end{observation}

  \tool{\texttt{goto 2830}}
  \begin{observation}
  File: root.ecore (13572 lines total)\\
  ...\\
  2828: <eClassifiers\\
  2829: \hspace{0.5cm} xsi:type="ecore:EClass"\\
  2830: \hspace{0.5cm} name="Frequency">\\
  ...
  \end{observation}

  \finalanswer{
  The Frequency class has the following fields:

  1. name: value \\
  \hspace*{0.5em} - type: EDouble (double precision floating point) \\
  \hspace*{0.5em} - default value: 1000 \\
  \hspace*{0.5em} - description: The numerical value of the quantity.

  2. name: unit \\
  \hspace*{0.5em} - type: FrequencyUnit (enumeration) \\
  \hspace*{0.5em} - default value: MHz \\
  \hspace*{0.5em} - description: The unit of the quantity.
  }
\end{minipage}
\end{tabular}

\caption{Example responses to a model-related question using (a) Direct Full-Context Prompting and (b) Agent with File Tools. The agent's directory-related tool call is omitted for clarity.}
\label{fig_example_responses}
\endgroup
\end{figure}

\subsubsection{Reference: Direct Full-Context Prompting}

The reference architecture represents a simple, baseline setup where the entire software model is placed directly into the LLM’s context window. The LLM is asked to answer questions in a zero-shot manner, with no examples or intermediate reasoning steps. This approach assumes the model fits into the context window of the evaluated LLM - an increasingly limiting factor for larger models, motivating the agentic approach.

The LLM prompt in this setup utilizes the Ecore format~\cite{noauthor_eclipse_nodate} for the input model, primarily because it's the native format of the tool used in our experiments. This choice streamlines the process by avoiding the need to translate the software model, even though formats like PlantUML or Mermaid~\cite{noauthor_mermaid_nodate} could also represent class relationships. Ecore, however, provides a more detailed and inherently tool-compatible representation. Consequently, this XML-based format requires more tokens to describe the software model than simpler, text-based syntaxes. Additionally, the prompt explicitly directs the LLM to analyze class relationships thoroughly and to rely exclusively on the provided content, assuming it is a complete and self-contained definition.

\subsubsection{ReAct Architecture: Agent with File Tools}

The second approach uses a tool-augmented agent architecture based on the ReAct framework~\cite{yao_react_2022}, where the LLM can choose to issue tool calls if it determines that the answer cannot be derived directly. The context window initially contains only the path to the directory with the software model files, not the model content itself. This setup allows the agent to selectively retrieve and reason over relevant parts of the model on demand.

The agent is equipped with a set of tools adapted from the SWE-agent framework~\cite{yang_swe-agent_2024}, which simulates an IDE-like interaction by exposing a windowed view over source files. Table~\ref{tab_file_tools} lists the tools available to the agent.

\begin{table*}[htbp]
\caption{File Tooling Capabilities in Agent-Based Architecture}
\centering
\begin{tabular}{|c|c|}
\hline
\textbf{Tool Name} & \textbf{Description} \\
\hline
\texttt{\detokenize{list_directory}} & Lists all files in a directory in a tree-style format. \\
\texttt{\detokenize{find_file}} & Finds all files with a given name or pattern. \\
\texttt{\detokenize{search_directory}} & Searches for a term across all files in the directory. \\
\texttt{\detokenize{open_file}} & Opens a file at a given path. Optionally scrolls to a specified line. \\
\texttt{\detokenize{scroll_up}} & Scrolls the open file window up by a fixed number of lines. \\
\texttt{\detokenize{scroll_down}} & Scrolls the open file window down by a fixed number of lines. \\
\texttt{\detokenize{go_to_line}} & Moves the window view to a specific line number in the open file. \\
\texttt{\detokenize{search_file}} & Searches for a term within the open file (or a specified file). \\
\hline
\end{tabular}
\label{tab_file_tools}
\end{table*}

Although some tools (\texttt{\detokenize{list_directory}}, \texttt{\detokenize{find_file}}, and \texttt{\detokenize{search_directory}}) are more relevant for models distributed across multiple files, they were retained for consistency and to enable the reuse of the agent with other software model datasets. For this experiment, all interactions involved a single file.

The windowing parameters for file tools were selected empirically to balance efficiency and context completeness. A window size of 50 lines with an overlap of 2 lines was used. This configuration proved effective for the selected software model, as most type definitions could be captured within one or two windows, thereby minimizing the need for extensive scrolling. At the same time, the window size was kept small enough to maintain low token usage and avoid exceeding the LLM’s context limits during multi-step reasoning.

The agent was allowed up to 100 iterations per question to avoid infinite loops while preserving enough flexibility for complex queries.

The agent prompt followed a similar structure to the reference setup, stating that the software model is in Ecore format and emphasizing that the model is complete.

\subsection{LLM Selection and Configuration}

This study evaluates LLMs that support tool usage - a capability essential for agent-based workflows. While direct prompting can, in principle, use any LLM, we employed the same tool-capable models for both approaches to ensure a fair comparison. In line with the privacy and deployment constraints typical in the automotive industry, we concentrated on smaller LLMs that are more feasible to run locally or on secure, private infrastructure. All selected models belong to the latest two generations of LLM development, offering enhanced reasoning capabilities and extended context lengths. We selected GPT-4o~mini, GPT-4.1~mini, and o4-mini to represent OpenAI’s current lightweight models with progressively improving architecture and performance characteristics, with o4-mini offering the strongest reasoning performance within this group. To broaden the evaluation, we also included Gemini 2.5 Flash Preview, a fast, tool-capable model from a different vendor, to assess cross-vendor performance and introduce architectural diversity. Table~\ref{tab_llms} summarizes the selected models.

\begin{table*}[htbp]
\caption{Language Models Used in the Experiments}
\centering
\begin{tabular}{|c|c|c|c|c|}
\hline
\textbf{Model Name} & \textbf{Release Date} & \textbf{Input Token Limit} & \textbf{Output Token Limit} & \textbf{Key Characteristics} \\
\hline
GPT-4o mini & 2024-07-18 & 128,000 & 16,384
& Compact multimodal model for efficient reasoning. \\
GPT-4.1 mini & 2025-04-14 & 1,047,576 & 32,768
& High-performance multimodal reasoning and coding. \\
o4-mini & 2025-04-16 & 200,000 & 100,000 
& Lightweight model for chain-of-thought reasoning. \\
Gemini 2.5 Flash Preview & 2025-04-17 & 1,048,576 & 65,536
& Fast, multilingual reasoning and summarization. \\
\hline
\end{tabular}
\label{tab_llms}
\end{table*}

In all agent configurations, the temperature parameter was set to 0.0 to ensure deterministic behavior and maximize answer reproducibility. The only exception is o4-mini, where the temperature is fixed at 1.0.

\subsection{Question Dataset Design}

As no existing datasets are tailored for evaluating LLMs on software model understanding, we constructed a small, custom dataset based on an open-source metamodel in Ecore format from INCHRON’s am2inc repository~\cite{noauthor_inchrons_nodate}. The am2inc tool facilitates the conversion of AMALTHEA models~\cite{iwai_eclipse_nodate} into the INCHRON format, which is then utilized by chronSUITE 3.X~\cite{noauthor_chronsuite_nodate}. AMALTHEA provides an open and unified model platform for multi-core software and hardware development. Its subsequent evolution into Eclipse APP4MC~\cite{iwai_eclipse_nodate} has significantly broadened its adoption, establishing it as an essential tool for analyzing and optimizing multi-core embedded systems, particularly within automotive applications. The am2inc conversion process is crucial because it filters out extraneous information irrelevant to timing analysis, thereby streamlining the model for chronSIM's specific focus on timing challenges~\cite{noauthor_inchrons_nodate-1}.

The selected metamodel root.ecore is substantial, defining 384 classes and consisting of 13,572 lines. When tokenized using the OpenAI tokenizer for GPT-4o, this translates to approximately 118,000 tokens. This token count highlights that while the model is considerably large and complex, it remains small enough to fit entirely within the context windows of modern LLMs - a crucial requirement for our experiments. Consequently, the selected INCHRON model serves as a representative example of real-world models used in the automotive industry for detailed timing analysis and optimization.

A total of 20 questions were manually created to cover different reasoning and retrieval patterns, grouped into four categories:

\begin{enumerate}
    \item Direct class inspection (5 questions): \\
Require identifying only the fields (names and types) of a specified class, without considering inheritance.

    \item Single inheritance chain (5 questions): \\
Require collecting fields from a class and its base classes along a single inheritance path.

    \item Multiple inheritance chains (5 questions): \\
Require traversing and aggregating fields across multiple inheritance paths.

    \item Semantic queries (5 questions): \\
Require understanding the model structure and semantics beyond keyword matching.
    
\end{enumerate}

This dataset enables controlled evaluation of both retrieval-based and reasoning-intensive tasks, simulating realistic software model comprehension needs in automotive software engineering workflows.

\subsection{Evaluation Criteria}

This study evaluates the performance of LLMs and agent-based architectures in answering questions about a large software model, using factual correctness as the primary metric. Factual correctness measures whether the statements in an LLM-generated answer are objectively true with respect to the software model~\cite{wang_factuality_2024}. It ensures that the output neither hallucinates, omits, nor misrepresents key information. In this context, the software model defines the complete scope of truth, including components, interfaces, dependencies, and structural or behavioral constructs described in the Ecore format.

Due to the number and structure of questions, as well as the number of evaluated setups, full human evaluation was deemed impractical. Instead, LLMs were used as proxies for scoring (LLM-as-a-Judge), with manual spot-checking performed for a randomly selected subset of answers to ensure reliability. Two complementary scoring methods were used:

\paragraph{Per-answer scoring}

Each generated answer was compared against a reference answer using a binary judgment, classified as either correct or incorrect. The accuracy metric was then computed as the ratio of correct answers to total questions. Evaluations were performed using GPT-4.1~mini with a structured prompt taken from the LangSmith Hub~\cite{noauthor_langsmith_nodate}. The output format was constrained to structured integer responses, where a score of 1 indicated correctness and 0 indicated incorrectness.

\paragraph{Per-fact scoring}

To capture more nuanced factual alignment, we employed the \texttt{FactualCorrectness} metric provided by Ragas version 0.2.15~\cite{noauthor_ragas_nodate}. This metric decomposes the model output and the reference into individual claims and applies natural language inference to identify factual overlap. The same LLM model, GPT-4.1~mini, is used here to perform the inference.

The evaluation returns precision, recall, and F1 scores, defined as follows:

\begin{align}
\text{Precision} &= \frac{TP}{TP + FP} \\
\text{Recall} &= \frac{TP}{TP + FN} \\
F_1\ \text{score} &= \frac{2 \cdot \text{Precision} \cdot \text{Recall}}{\text{Precision} + \text{Recall}}
\end{align}
where true positives (TP) refer to claims in the response that also appear in the reference. False positives (FP) are claims present in the response but not in the reference, while false negatives (FN) are claims in the reference that are missing from the response.

This scoring method enables more granular insight into partially correct responses and helps differentiate between omission and hallucination errors.

\section{Results and Analysis}


This section presents the experimental results of direct prompting and agent-based LLM architectures for querying software models. We evaluated these methods using multiple LLM variants on a 20-question dataset, focusing on answer correctness and token efficiency.

\subsection{Correctness Evaluation}

Table~\ref{tab_factual_correctness_results} presents the factual correctness results of all evaluated architectures and LLMs. Overall, the direct full-context prompting setup performed better or comparably for several models. Notably, Gemini 2.5 Flash and GPT-4o~mini achieved significantly higher accuracy in the reference setup compared to the agent-based configuration. Gemini 2.5 Flash reached 80\% accuracy in the direct prompting, but only 40\% when operating within the ReAct-based tool-augmented agent. Similarly, GPT-4o~mini's performance dropped sharply from 45\% to 10\%. This suggests that these models may struggle with tool-based reasoning, maintaining coherence across tool calls, or operating effectively within constrained, windowed views of the software model.

In contrast, GPT-4.1~mini and o4-mini maintained high accuracy (90\%) across both setups. Their performance was especially strong in the agent with file tools architecture, indicating better capabilities for orchestrating tool use and reasoning over partial views of the model. These results support the hypothesis that agent-based setups place greater demands on a model’s reasoning and memory capabilities, which smaller or less capable LLMs may not handle well.

Another key observation is the precision-recall trade-off visible across models. GPT-4.1~mini in the direct prompting, for instance, achieved high recall (0.86) but low precision (0.44), implying a tendency to include more relevant information at the cost of introducing inaccuracies.

Finally, the relatively high standard deviations in precision, recall, and F1 score across models (up to $\pm$0.36) suggest considerable variability in model performance depending on the specific question type. This highlights the importance of comprehensive evaluation across diverse question categories when assessing LLM effectiveness in model understanding tasks.

\begin{table*}[htbp]
\caption{Factual correctness across 20 questions for evaluated setups: Accuracy (\%), Precision, Recall, F1 (mean $\pm$ std).}
\centering
\begin{tabular}{|c|c|c|c|c|c|c|c|}
\hline
\textbf{Architecture} & \textbf{LLM Model} & \textbf{Correct} & \textbf{Incorrect} & \textbf{Accuracy (\%)} & \textbf{Precision ($\pm$SD)} & \textbf{Recall ($\pm$SD)} & \textbf{F1 score ($\pm$SD)} \\
\hline
\multirow{4}{*}{\makecell{Direct\\Full-Context\\Prompting}} 
& Gemini 2.5 Flash & 16 & 4  & 80.0             & \underline{0.70 $\pm$ 0.29} & \underline{0.91 $\pm$ 0.25} & \underline{0.75 $\pm$ 0.27}  \\
& GPT-4.1 mini     & 16 & 4  & 80.0             & 0.44 $\pm$ 0.23             & 0.86 $\pm$ 0.30      
      & 0.56 $\pm$ 0.23  \\
& GPT-4o mini      & 9  & 11 & 45.0             & 0.53 $\pm$ 0.23             & 0.79 $\pm$ 0.28      
      & 0.61 $\pm$ 0.22  \\
& o4-mini          & 18 & 2  & \underline{90.0} & 0.59 $\pm$ 0.23             & 0.90 $\pm$ 0.24      
      & 0.66 $\pm$ 0.22  \\
\hline
\multirow{4}{*}{\makecell{Agent with\\File Tools}} 
& Gemini 2.5 Flash & 8  & 12 & 40.0             & \underline{0.59 $\pm$ 0.36} & 0.68 $\pm$ 0.35 & 0.58 $\pm$ 0.36  \\
& GPT-4.1 mini     & 18 & 2  & \underline{90.0} & 0.49 $\pm$ 0.23             & \underline{0.95 $\pm$ 0.11} & \underline{0.61 $\pm$ 0.19}  \\
& GPT-4o mini      & 2  & 18 & 10.0             & 0.19 $\pm$ 0.31             & 0.21 $\pm$ 0.35      
      & 0.20 $\pm$ 0.32  \\
& o4-mini          & 18 & 2  & \underline{90.0} & 0.48 $\pm$ 0.36             & 0.85 $\pm$ 0.37      
      & 0.57 $\pm$ 0.35  \\
\hline
\end{tabular}
\label{tab_factual_correctness_results}
\end{table*}

\subsection{Token Usage Comparison}

To better understand the computational footprint and processing patterns of the evaluated setups, we analyzed token usage across all models and architectures. The results are summarized in Table~\ref{tab_token_count_results}.

A striking, though not surprising, difference is observed in the number of prompt tokens between the direct full-context prompting and the agent with file tools. Direct prompting relies on injecting the entire software model directly into the prompt, resulting in high prompt token counts - typically ranging from 118,000 to over 137,000 tokens on average, depending on the tokenizer used. In contrast, the agent setup loads context incrementally via file tools, resulting in dramatically lower prompt usage—only about 640–780 tokens depending on the model. This demonstrates that agent-based designs are far more efficient in managing input context.

In terms of completion tokens, which represent the size of the final answer, the values are consistently small across all setups. However, they are slightly higher in the reference setup, likely because the model has access to the entire software model upfront and can produce more detailed responses. The agent-based setups produce shorter completions, possibly due to more focused context windows available at response time.

Reasoning token count offers insight into how models internally process queries. GPT-4.1 mini and GPT-4o mini show no reasoning tokens, suggesting they produce answers without explicit intermediate steps. In contrast, Gemini 2.5 Flash and o4-mini exhibit substantial built-in reasoning. Both models use significantly fewer reasoning tokens in the agent configuration than in direct prompting, likely due to the modular, tool-driven context limiting the need for extended internal processing.

Overall, these findings confirm that agent-based approaches are significantly more efficient in token usage, especially for large software models, while still enabling internal reasoning in models designed for that purpose. They also highlight that reasoning behaviors vary greatly depending on model capabilities.

\begin{table*}[htbp]
\caption{Token usage across evaluated setups. Mean values with standard deviations are reported.}
\centering
\begin{tabular}{|c|c|c|c|c|c|}
\hline
\textbf{Architecture} & \textbf{LLM Model} & \textbf{Prompt Tokens ($\pm$SD)} & \textbf{Completion Tokens ($\pm$SD)} & \textbf{Reasoning Tokens ($\pm$SD)} & \textbf{Total Tokens ($\pm$SD)} \\
\hline
\multirow{4}{*}{\makecell{Direct\\Full-Context\\Prompting}} 
& Gemini 2.5 Flash & 137461.4 $\pm$ 6.1 & 174.7 $\pm$ 138.8 & 598.0 $\pm$ 328.8 & 138234.1 $\pm$ 404.4  \\
& GPT-4.1 mini     & 118776.0 $\pm$ 5.8 & 210.0 $\pm$ 129.4 & 0.0 $\pm$ 0.0     & 118986.0 $\pm$ 131.5  \\
& GPT-4o mini      & 118776.0 $\pm$ 5.8 & 223.7 $\pm$ 164.0 & 0.0 $\pm$ 0.0     & 118999.7 $\pm$ 166.5  \\
& o4-mini          & 118775.0 $\pm$ 5.8 & 150.3 $\pm$ 95.1  & 563.2 $\pm$ 291.9 & 119488.5 $\pm$ 355.2  \\
\hline
\multirow{4}{*}{\makecell{Agent with\\File Tools}} 
& Gemini 2.5 Flash & 783.3 $\pm$ 6.1 & 17.6 $\pm$ 6.5 & 282.5 $\pm$ 60.6 & 1083.4 $\pm$ 65.5  \\
& GPT-4.1 mini     & 639.3 $\pm$ 5.7 & 17.7 $\pm$ 1.4 & 0.0 $\pm$ 0.0    & 656.9 $\pm$ 5.8  \\
& GPT-4o mini      & 642.6 $\pm$ 2.1 & 30.9 $\pm$ 4.3 & 0.0 $\pm$ 0.0    & 673.4 $\pm$ 3.6  \\
& o4-mini          & 641.8 $\pm$ 6.0 & 53.4 $\pm$ 9.4 & 60.4 $\pm$ 40.9  & 755.7 $\pm$ 45.4  \\
\hline
\end{tabular}
\label{tab_token_count_results}
\end{table*}

\subsection{Failure Case Analysis}

Two main categories of failures were observed during the evaluation: agent errors, which prevented the model from producing an answer, and incorrect answers, where the model made a factually inaccurate response.

The only error type observed in the agent-based architecture was a recursive error, occurring when the number of tool-use iterations exceeded the predefined recursion limit of 100. These errors prevented the agent from producing any answer, leading to a failed attempt. Across all evaluations, recursive errors occurred 15 times when using the agent with file tools. Two of these cases were attributed to the o4-mini model. In both instances, the agent repeatedly opened the same file and interleaved file searches with opening the file at different lines, continuing this behavior until the iteration limit was reached. Some search terms appeared arbitrary or incorrectly chosen, which may point to a misunderstanding of Ecore syntax. The remaining 13 failures were observed with GPT-4o~mini. In twelve of these, the agent opened a file and continuously scrolled down until it hit the iteration cap. In one case, the agent alternated between scrolling down and up while occasionally invoking \texttt{\detokenize{find_file}} with class names, again without arriving at a coherent answer. All such recursive errors were treated as incorrect responses and were therefore assigned a score of 0.0 in the precision, recall, and F1 score metrics.

The main observation from the incorrect answers is that the dominant failure mode across models - especially in tasks involving class field and inheritance chain extraction - is incomplete or prematurely terminated inheritance traversal. Many agents stop after one or two inheritance levels, failing to follow the full chain and include all inherited fields, which leads to missing attributes. Another frequent issue is confusion or inconsistency in attributing fields to the correct base class, often misplacing a field’s origin in the hierarchy. Some answers are penalized due to minor factual inaccuracies, such as incorrect data types or multiplicity (e.g., treating a field as a list where it’s not explicitly typed as such in the model). In cases where class membership questions are involved (e.g., bus ports, event chain requirements), models often overgeneralize by including additional classes with semantically similar names or nearby definitions, which, while arguably plausible, deviate from the ground truth. Notably, models using file tools often exhibit partial exploration behaviors, such as navigating to a class but not scrolling to see all fields, or identifying a base class but not recursively inspecting further base classes up the inheritance chain. These patterns suggest that incorrect answers often arise not from fundamental misunderstanding but from shallow or incomplete retrieval and reasoning processes.

\section{Discussion}

Our results reveal several essential insights. First, the reference architecture generally yields higher or comparable factual correctness for most models when the full model fits into the LLM’s context window. This outcome is intuitive since the model has complete visibility of the entire software structure at once, enabling holistic reasoning and direct fact retrieval. Notably, models such as Gemini 2.5 Flash and GPT-4o~mini exhibited a significant drop in accuracy when constrained to the agent setup. This suggests that certain LLM variants may struggle with maintaining state coherence across multiple tool calls or with reasoning effectively over fragmented, partial views of the model. Moreover, analysis of incorrect answers shows that the models frequently failed to choose the most effective tool (\texttt{\detokenize{search_file}}) for the task, instead relying on a less efficient option (\texttt{\detokenize{scroll_down}}).

In contrast, GPT-4.1~mini and o4-mini maintained robust accuracy (90\%) across both architectures, highlighting the importance of advanced reasoning and memory capabilities for effective tool-assisted model comprehension. Their success in the agent setup demonstrates that an LLM capable of efficiently orchestrating the retrieval and synthesis of partial information can overcome the limitations imposed by lacking full context. However, analysis of incorrect answers suggests that accuracy could be further improved by integrating tools specifically designed for browsing software models, which better address the challenges of partial exploration, rather than relying solely on general-purpose file access tools. This insight has important practical implications for industrial workflows, where software models often exceed the size limits of even the largest available context windows.

The observed precision-recall trade-offs provide important insights. As expected, high recall generally corresponds to higher overall accuracy in most cases, indicating that models tend to capture relevant information effectively. However, overall precision remains relatively low across models, suggesting a common tendency to include extra or occasionally inaccurate details. Striking the right balance between recall and precision is especially critical in engineering domains, where accuracy is paramount and any hallucinations or errors can mislead developers or compromise system safety.

From an efficiency perspective, agent-based architectures dramatically reduce prompt token usage by avoiding full ingestion of large software models. This efficiency gain is crucial for enabling deployment of LLMs with smaller context windows on-premise, addressing automotive industry constraints around privacy, intellectual property protection, and regulatory compliance. Although completion tokens (the size of answers) were similar across setups, the reduction in prompt size directly translates into lower computational costs, faster response times, and the feasibility of more frequent or iterative querying.

\section{Threats to Validity}

\paragraph{Single Model and Format}
The study used a single software model in Ecore format, limited to a single file. This constrains generalizability, as other models, formats, or multi-file setups may present different challenges.

\paragraph{Limited and Narrow Question Scope}
The evaluation was based on 20 questions focused solely on structural, low-level aspects (e.g., class fields). It did not assess behavioral properties or higher-abstraction queries.

\paragraph{Answer Validation Limitations}
Extra statements generated by the LLM beyond the expected answers were not verified for correctness, which may have impacted precision and recall metrics. Additionally, the evaluation relied on LLM-based scoring methods, potentially introducing inaccuracies in answer correctness determination.

\paragraph{Fixed Agent Parameters}
The agent used a fixed configuration (e.g., window size, iteration limits), which may have limited adaptability or performance across different tasks.

\section{Conclusions and Future Work}

This study compared direct prompting and agent-based approaches for answering questions over software models using LLMs. The agentic approach proved effective with smaller LLMs, demonstrating that compact models possess strong tool usage and reasoning capabilities. This improvement makes sophisticated, retrieval-augmented agent workflows feasible on resource-constrained setups. The results highlight the potential of such structured approaches for interacting with large software artifacts efficiently and accurately.

Future work will explore several directions. First, the approaches will be evaluated in a broader variety of software models and formats to assess generalizability. More complex agent architectures will be investigated to improve reasoning capabilities and task coordination, leveraging emerging standards such as MCP and A2A to enhance tool integration and agent collaboration. The question set will be expanded, including queries at different abstraction levels to probe deeper understanding. Additionally, efforts will be made to automatically generate model instances from metamodels and specifications, and to support textual modifications of software models while preserving consistency.

\bibliographystyle{IEEEtran}
\bibliography{references_abrv}

\end{document}